\documentclass{aipproc}
\usepackage{graphicx}

\layoutstyle{6x9}
\begin{document}

\vspace*{0.5cm}
\title{The pion-pion scattering lengths from DIRAC}
\XFMtitle

\vspace*{-1ex}
\begin{center}
C.~Santamarina \\
{\it Basel University, Switzerland}

\smallskip
on behalf of DIRAC Collaboration\\
\end{center}

\vspace*{-1ex}

\begin{abstract}
  The scattering lengths of a two pion system are the {\it golden
  magnitudes} to test the QCD predictions in the low energy sector.
  The DIRAC (PS-212) experiment at CERN will obtain a particular
  combination of the S-wave isospin 0 and 2 scattering lengths by
  measuring the lifetime of pionium, the hydrogen-like $\pi^+ \pi^-$
  atom. This measurement tests the accurate predictions of the Chiral
  Perturbation Theory. The most recent experimental results are presented.
\end{abstract}

\XFMabstract

%\maketitle

\section{Introduction}%1

DIRAC experiment, conducted at CERN, is
measuring the lifetime of the ground state of pionium, the hydrogen-like
$\pi^+ \pi^-$ atom~\cite{prop}. The decay of pionium is dominated by the
strong channel $\pi^+ \pi^- \rightarrow \pi^0 \pi^0$ (BR=$94\%$). The 
ground state lifetime is related to the S-wave isospin 0 and 2 scattering
length difference by the Deser type formula known to NLO~\cite{gall}:
\begin{equation}
\frac{1}{\tau} = \frac{2}{9} \alpha^3 \sqrt{M_{\pi}^2-M_{\pi^0}^2
-\alpha^2 M_{\pi}^2/4} \left|a_0^0-a_0^2\right|^2 (1+\delta),
\end{equation}
where $\alpha$ is the fine structure constant, $M_{\pi}$ the mass of the
charged pions, $M_{\pi^0}$ the mass of the neutral pion and
$\delta$ the correction to NLO ($\delta = 0.058$).

The Chiral Perturbation Theory has a precise prediction to $\mathcal{O}(p^6)$
for this difference and hence for the lifetime of pionium~\cite{cola}:
\[
\left| a_0^0-a_0^2 \right| = 0.265 \pm 0.004 \Rightarrow 
\tau = (2.9\pm 0.1)\cdot 10^{-15} s,
\]
which must be checked with similarly accurate experimental data. DIRAC goal is
to achieve a $5\%$ precision in the determination of the scattering length
difference by measuring the lifetime with $10\%$ precision.

\section{Experimental Method}%2

DIRAC measures the pionium lifetime analyzing
the low relative momentum ($\vec{Q}$) spectrum of $\pi^+ \pi^-$ pairs
produced in the collisions of the $24 GeV/c$ 
proton beam with the target. The target consists of one or several metal layers
of the overall thickness of $\sim 100 \mu m$.

\begin{figure}
\includegraphics[width=0.8\textwidth]{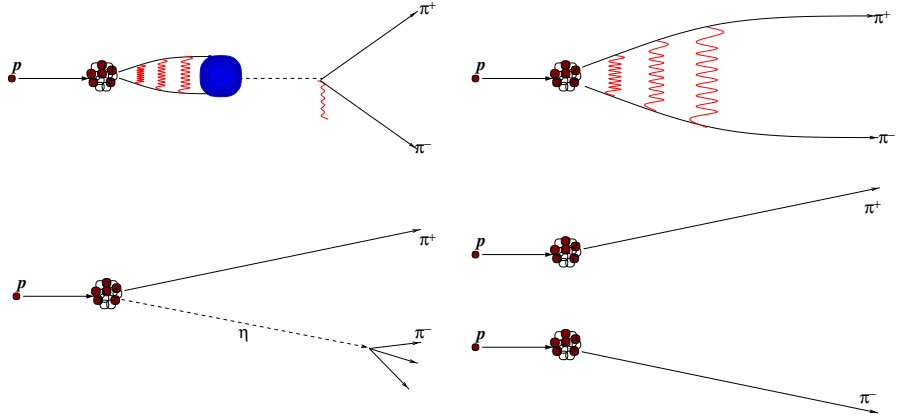}
\caption{Pion pair production mechanisms: {\bf Top left:}{\em Atomic Pair}.
{\bf Top right:}{\em Coulomb Pair}. {\bf Bottom left:}{\it Non Coulomb Pair}.
{\bf Bottom right:} {\it Accidental Pair}.}
\label{scheme}
\end{figure}

This spectrum is formed of {\it Real} (time correlated) pairs, in
which both pions
come from a single proton-target interaction, and {\it Accidentals}, coming
from two different interactions. Time correlated pairs are divided into:
\begin{itemize}
\item {\it Atomic Pairs (pairs resulting from atomic break-ups)}. Pionic atoms are created
by the Final State Coulomb Interaction (FSCI) of low relative momentum
$\pi^+ \pi^-$ pairs.
The sample of initial pionic atoms ($N^A$) evolves
by colliding with the
atoms of the metal target. The evolution terminates when the atoms
break-up in one of the collisions or annihilate. If the number of
broken atoms is given by $n^A$ then, the shorter is
the lifetime of pionium the smaller is the break-up
probability $P_{br}=n^A/N^A$. The dependence
of the break-up probability on the lifetime is accurately known~\cite{pbrr}, as shown
in figure~\ref{ff} for the Nickel target. {\it Atomic Pairs} are restricted to the $Q<4 MeV/c$ region.

\item {\it Coulomb Pairs}. Their production mechanism is the same
as for the
atoms but leading to continuous spectrum eigenstates of the hydrogen-like
hamiltonian of the $\pi^+ \pi^-$ system. This is why the yield of
atoms is proportional to the number of Coulomb pairs within a determined
$\vec{Q}\in \Omega$ arbitrary region~\footnote{In the experiment usually
$\Omega = \{ \mbox{Events with reconstructed $Q<2 MeV/c$}\}$.}~\cite{neme}:
\begin{equation}
N^A = K \times \int_{\vec{Q}\in \Omega} \frac{dN^C}{dQ}.
\end{equation}
The relative momentum distribution of {\it Coulomb Pairs} is given by:
\begin{equation}
\frac{dN^C}{dQ} \propto \mathcal{T}\left[ \frac{2\pi M_{\pi}\alpha/Q}
{1-e^{-2\pi M_{\pi}\alpha/Q}} Q^{2} \right],
\end{equation}
where $\mathcal{T}$ stands for the transformation due to multiple scattering in the
target and setup resolution of the {\it at production} distribution between brackets.
\item {\it Non Coulomb Pairs}. Non Correlated pairs in which
one of the pions comes from the decay of a long-lived source~\footnote{A long
lived source is a resonance with decay time large enough that it can fly
a distance much larger than the pionium Bohr radius ($387 fm$) from the primary
interaction vertex.}.
Its relative momentum distribution is purely phase space driven and, hence, given by
\begin{equation}
\frac{dN^{NC}}{dQ}  \propto \mathcal{T}\left[Q^2\right].
\end{equation}
\end{itemize}

Apart from time correlated pairs also {\it Accidental} $\pi^+ \pi^-$ {\em Pairs},
where each pion comes from a different proton-target interaction, are recorded.
Their relative momentum distribution is also determined by phase space:
\begin{equation}
\frac{dN^{Acc}}{dQ}  \propto \mathcal{T}\left[Q^2\right].
\end{equation}
The {\it Accidental Pairs} sample is used to parameterize the
{\em Real Pairs} pairs spectrum:
\begin{equation}
\frac{dN^{Real}}{dQ} = \frac{dn^A}{dQ} + \frac{dN^{NC}}{dQ}
+ \frac{dN^C}{dQ} = \frac{dn^A}{dQ} + (a R(Q)+b) \frac{dN^{Acc}}{dQ},
\end{equation}
where $R(Q)$ is the ratio between Coulomb pairs and accidentals, obtained with
Monte Carlo, and $a$ and $b$ are calculated by fitting the spectrum in the
$Q>4MeV/c$ region, free from atomic pairs.

The shape of the relative momentum distribution of {\it Atomic Pairs} 
can be obtained as:
\begin{equation}
\label{atomic}
\frac{dn^A}{dQ} = \frac{dN^{Real}}{dQ} - (a R(Q)+b) \frac{dN^{Acc}}{dQ}.
\end{equation}
Moreover, the break-probability can then be experimentally determined:
\begin{equation}
%P_{br} =  \frac{\int \frac{dn^A}{dQ}}
%{K \int_{\vec{Q}\in \Omega} a R(Q) \frac{dN^{Acc}}{dQ}}.
P_{br} =  \int \frac{dn^A}{dQ} \left/
K \int_{\vec{Q}\in \Omega} a R(Q) \frac{dN^{Acc}}{dQ} \right. .
\end{equation}

An alternative method consists of using pure Monte Carlo distributions for {\it Coulomb},
{\it Non Coulomb} and {\it Atomic Pairs} to parameterize the {\it Real}
pairs spectrum:
\begin{equation}
\frac{dN^{Real}}{dQ} = \alpha \frac{dn^A_{MC}}{dQ} + 
\beta \frac{dN^{NC}_{MC}}{dQ}
+ \gamma \frac{dN^C_{MC}}{dQ},
\end{equation}
$\alpha$, $\beta$ and $\gamma$ being parameters obtained from the fit
to the whole spectrum~\cite{schuetz}.

%The data analysis can be also performed on $Q_L$, the longitudinal
%component of the relative momentum, which is less sensitive to the
%multiple scattering in the target but has the disadvantage that the number
%of atomic paris int the low $Q$ region is only a very small percentage
%and leading to larger statistical errors.

\section{Spectrometer}%3

DIRAC spectrometer is a two arm telescope with a 1.65 $T$ central magnet
separating positive and negative charged particle channels~\cite{dirac}. It is located in the T8 channel
of the PS East Hall B at CERN. The beam intensity is $\sim 10^{11}$ 24 GeV/c protons per spill (0.4 s).

The detectors are classified into {\it upstream}
and {\it downstream} ones according to their position relative to the magnet.
The {\it upstream} detectors consist of three sets of planes of Micro Strip Gas Chambers
and Scintillating Fibers, used in tracking, and Ionization Hodoscopes, also used for triggering.
{\it Downstream} we have two sets of Drift Chambers, the main tracking
detector, Vertical and Horizontal hodoscope arrays for timing and triggering, \v{C}erenkov
detectors, to reject electrons at trigger level, Preshower, also used for triggering and 
electrons rejection offline, and muon detectors, for muons rejection.

The trigger system consists of three levels~\cite{trig}. The T1 main trigger
selects $\pi^+ \pi^-$ events with coincidences
upstream and the two arms downstream. The DNA+RNA makes a topological on-line analysis
to perform the relative momentum cuts $Q_x < 3 MeV/c$, $Q_y < 10 MeV/c$ and $Q_l < 30 MeV/c$.
Finally, the T4 level selects events 
containing one track per arm satisfying $Q_x<3 MeV/c$ and $Q_l < 30 MeV/c$ criteria.

The track resolution uses the standard Kalman filter procedure achieving
an excellent relative momentum resolution of $\sigma_{Q_x}=\sigma_{Q_y}=0.4 MeV/c$ and
$\sigma_{Q_L}=0.6 MeV/c$.

\begin{figure}
\includegraphics[width=0.75\textwidth]{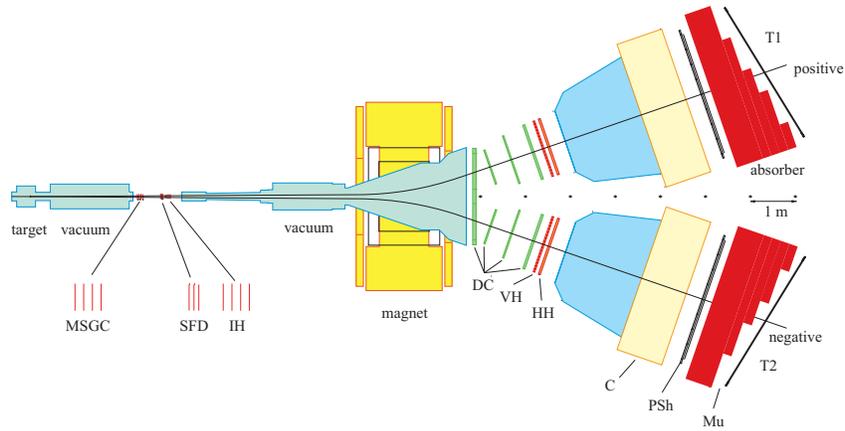}
\caption{DIRAC spectrometer scheme. The top view shows the distribution of the detectors.
MSGC=Micro Strip Gas Chambers, SFD=Scintillating Fiber Detector, IH=Ionization Hodoscopes detector,
DC=Drift Chambers, VH=Vertical Hodoscopes, HH=Horizontal Hodospes, C=\v{C}erenkov, PSh=Preshower,
Mu=Muon counters.}
\label{spectrometer}
\end{figure}

\section{Experimental results}%4

DIRAC started taking data in November 1998. Since that time we have collected
thousands of $\pi^+ \pi^-$ {\it Atomic Pairs}. For example, the accumulated Nickel target~\footnote{DIRAC
main target is made of Nickel. The former $94 \mu m$ thickness target was replaced in September 2001 by
a $98 \mu m$ one.}
statistics of 2000, 2001 and $75\%$ of 2003 runs with tight cuts gives more than 12000 pairs.

However, the accuracy of the measurement, has been limited by systematic effects as the behavior
of the electronic readout of some detectors or the parameterization of Multiple Scattering.
In the last two years the major thrust of our research has been the study of systematics,
the development of the Monte Carlo simulation of the experiment as well
as dedicated measurements like the use of a multi-layer target~\cite{mlayer} to deep in the knowledge of our data.
These studies are foreseen
to be finished early 2004. At the moment we have only a rough estimate of the systematic error of lifetime
$\sim 1\cdot 10^{-15} s$.
Accounting for that, the result obtained with the use of Nickel target 2001 
statistics, the best understood data, gives:
\[ \tau = (3.1^{+0.9}_{-0.7} (stat.) \pm 1. (syst)) \cdot 10^{-15} s. \]
The pure Monte Carlo analysis leads to a compatible result of
\[ \tau = (2.6 \pm 0.5 (stat.) \pm 1. (syst)) \cdot 10^{-15} s. \]

\begin{figure}
\includegraphics[width=0.8\textwidth]{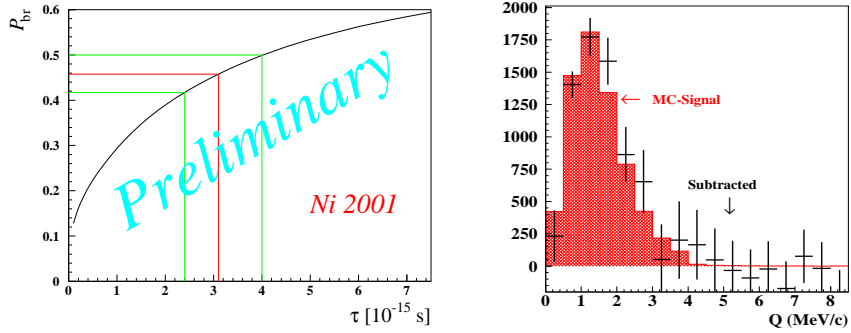}
\caption{{\bf Left:} Lifetime measurement. {\bf Right:} Measured spectrum
of $\pi^+ \pi^-$ {\it Atomic Pairs} and comparison to
the Monte Carlo corresponding distribution.}
\label{ff}
\end{figure}

In figure~\ref{ff} we can also see the experimental shape of the spectrum of
pairs from atomic break-up as defined in equation (\ref{atomic}).

\end{document}